\documentclass[aps,prl,twocolumn,groupedaddress]{revtex4}

\usepackage{graphicx} 
\usepackage{dcolumn}
\usepackage{bm}

\begin{document}
\title{Unusual temperature dependence of the oxygen-isotope effect on the exchange-energy of La$_{1-x}$Ca$_{x}$MnO$_{3}$
at high temperatures} 
\author{Guo-meng Zhao$^{*}$ and John Mann} 
\affiliation{Department of Physics and Astronomy, 
California State University, Los Angeles, CA 90032, USA}

\begin{abstract}
We report magnetic susceptibility $\chi(T)$ measurements on oxygen-isotope 
exchanged La$_{1-x}$Ca$_{x}$MnO$_{3+y}$ up to 700 K. The
$1/\chi(T)$ data show that the ferromagnetic exchange-energy $J$ depends strongly on the
oxygen-isotope mass. The isotope effect on $J$ decreases with temperature 
up to 400 K and then increases again with temperature above
400 K. This unusual temperature dependence of the isotope effect cannot be explained by existing 
theories of the colossal magnetoresistance effect for doped manganites.
The present results thus provide essential constraints on the physics
of manganites.
 
\end{abstract}
\maketitle

Doped manganites such as La$_{1-x}$Ca$_{x}$MnO$_{3+y}$ ($x$ $>$ 0.20) show a 
colossal magnetoresistance (CMR) effect \cite{Von}. This novel physical property makes them potentially important in technological applications. 
The understanding of the microscopic mechanism responsible for CMR has become one of the outstanding problems in condensed matter physics 
because no consensus concerning the CMR mechanism has been reached
\cite{Millis1,Roder,Moreo,Alex} despite tremendous theoretical and experimental efforts for 
over 10 years. Some outstanding problems of particular interest
include the observations of intrinsic inhomogeneity
\cite{Teresa,Uehara}, the giant oxygen-isotope effect on 
the Curie temperature ($T_{C}$) \cite{ZhaoNature,ZhaoPRL}, and the transition from a ferromagnetic metallic ground state to an antiferromagnetic charge-ordered 
insulating ground state by simply replacing $^{16}$O with $^{18}$O isotope
\cite{ZhaoPS,Ibarra}. These novel isotope effects have never been predicted by 
any conventional theories based on the Migdal approximation. The fact that lattice vibrations can significantly modify 
the electrical and magnetic properties of doped manganites suggests that the electron-phonon interactions are so strong that 
electronic and lattice subsystems are no longer decoupled. 

The physics of manganites 
has primarily been described by the
double-exchange (DE) model \cite{Zener}. However, Millis, Littlewood and
Shraiman (MLS) \cite{Millis1} pointed out that the carrier-spin 
interaction in the DE model is too weak to lead to the carrier 
localization in the paramagnetic state, and thus a second mechanism such as 
a small polaronic effect
is needed to explain the observed resistivity data 
in doped manganites. The central point of the model is that in the paramagnetic 
state the electron-phonon coupling constant $\lambda$ is large enough 
to form small polarons while the growing ferromagnetic order increases 
the bandwidth and thus decreases $\lambda$ sufficently to form 
a large polaron metallic state. Many experiments
\cite{ZhaoNature,ZhaoPRL,Jaime,Billinge,Teresa,Booth,Louca} have 
provided strong evidence for the 
existence of small polarons in the paramagnetic state, and 
qualitatively support the MLS and related models \cite{Millis1,Roder,Moreo}.
On the other hand, Alexandrov and Bratkovsky (AB) \cite{Alex} argue that the 
MSL model cannot quantitatively 
explain CMR, and thus propose an 
alternative CMR theory. The basic idea of their model is that the small polarons 
form localized bound pairs (bipolarons) in the paramagnetic state 
while the competing exchange interaction 
of polaronic carriers with localized spins drives the ferromagnetic 
transition. The transition is accompanied by a giant increase in the 
number of small polarons which are mobile carriers and move coherently 
at low temperatures. The AB model is consistent with the observed oxygen-isotope effect on the
intrinsic resistivity in the ferromagnetic state \cite{ZhaoPRBLT} and the temperature
dependence of the resistivity at low temperatures \cite{ZhaoPRL00}.  
Moreover, the observed oxygen-isotope effects on the intrinsic
resistivity and thermoelectric power in the paramagnetic state
\cite{ZhaoPRBHT} appear to be well explained by
the existence of localized bipolaronic charge carriers, which also 
supports the AB model.

Here we report magnetic susceptibility measurements on oxygen-isotope 
exchanged La$_{1-x}$Ca$_{x}$MnO$_{3+y}$ ($x$ = 0.25
and 0.33) up to 700 K. The
data of the inverse paramagnetic
susceptibility $1/\chi(T)$ show that the ferromagnetic exchange energy $J$ depends strongly on the
oxygen-isotope mass. The isotope effect on $J$ decreases with temperature 
up to about 400 K and then increases again with temperature above
400 K. This unusual temperature dependence of the isotope effect cannot be explained by both
AB and MLS models.

Samples of  La$_{1-x}$Ca$_{x}$MnO$_{3}$ were
prepared by
conventional solid state reaction using dried La$_{2}$O$_{3}$, MnO$_{2}$ 
and CaCO$_{3}$.  The well-ground mixture was 
calcined in air at 1000 $^{\circ}$C for 20 hours, 1100 $^{\circ}$C for 20
hours 
with one intermediate grinding. The powder 
samples were then pressed into pellets
and sintered  at 1260 $^{\circ}$C for 72 hours, and 1160 $^{\circ}$C for 
72 hours with one intermediate grinding. Two  pieces were 
cut
from the same pellet for oxygen-isotope diffusion. The diffusion was 
carried out
for 50 hours
at 1000 $^{\circ}$C and in oxygen partial pressure of about 1 bar. The cooling 
rate
was 300 $^{\circ}$C/hour.
The oxygen-isotope enrichment was determined from the weight changes 
of
both $^{16}$O and $^{18}$O samples. The $^{18}$O samples had 
$\sim$90$\%$
$^{18}$O
and $\sim$10$\%$ $^{16}$O. 

\begin{figure}[htb]
    \includegraphics[height=5.5cm]{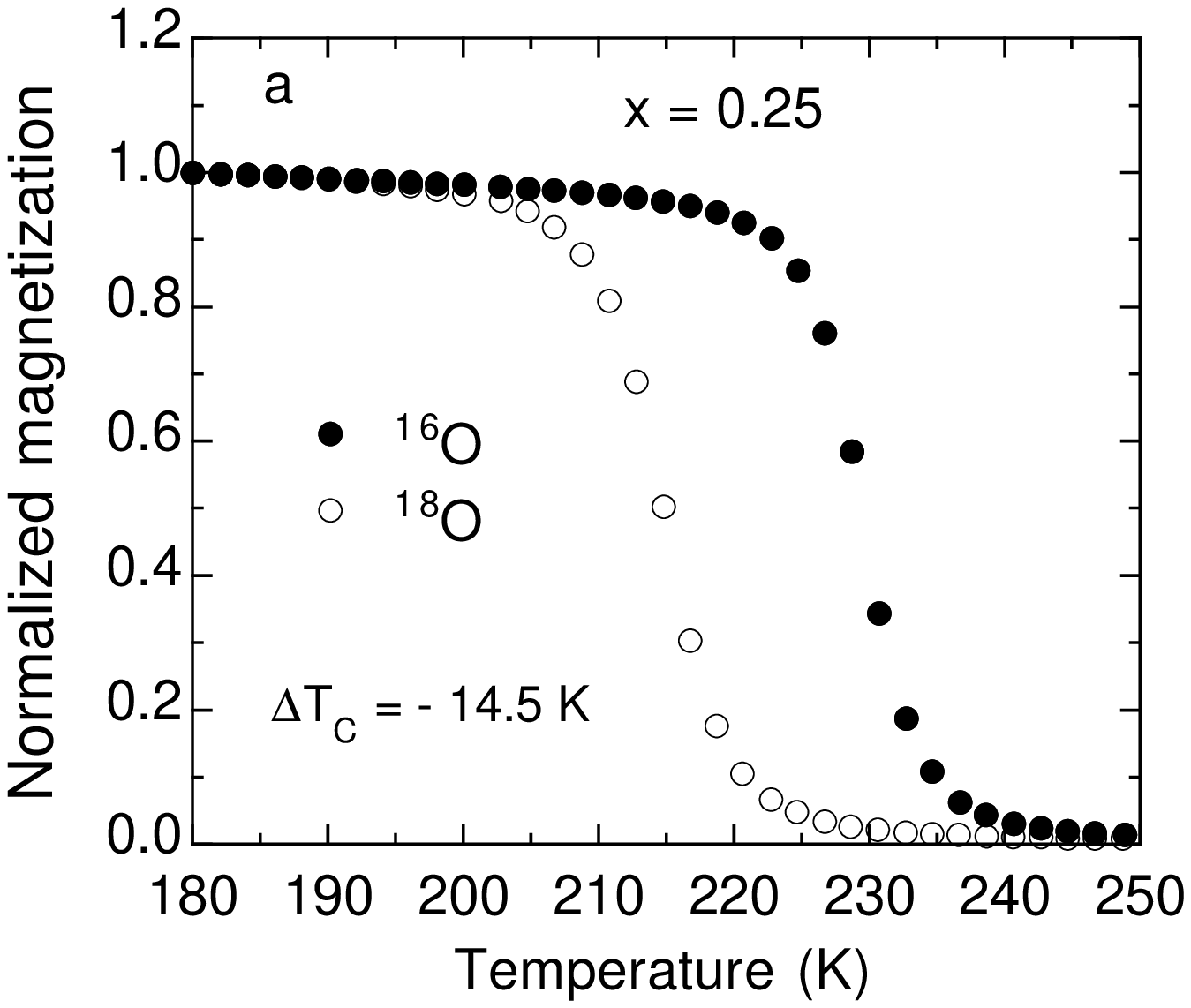}
	 \includegraphics[height=5.5cm]{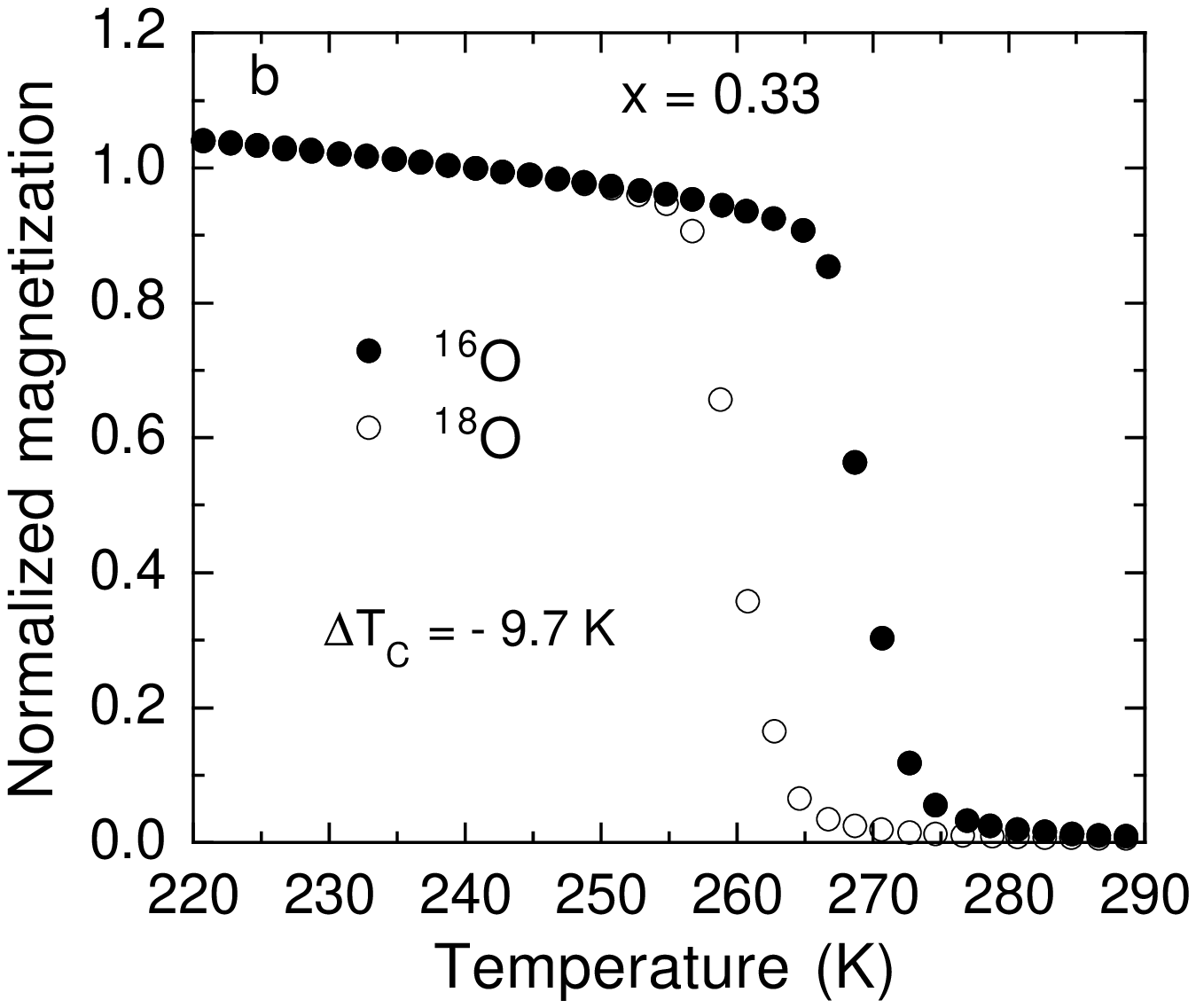}
 \caption[~]{Temperature dependencies of the normalized
magnetizations (normalized to the magnetization
well below T$_{C}$) for the $^{16}$O and $^{18}$O samples
of La$_{1-x}$Ca$_{x}$MnO$_{3}$ with  $x$ = 0.25 (a)
and $x$ = 0.33 (b).   }
\end{figure}

 Field-cooled magnetization below 300 K was measured with a Quantum Design SQUID 
 magnetometer in a field of  50 Oe.  The samples were cooled
directly to 5 K, then warmed up to a temperature well below T$_{C}$. 
After waiting for 5 minutes at that temperature, data were collected upon warming to a
temperature well above T$_{C}$. Magnetization between 300 K and 700 K
was measured with a Quantum Design vibrating sample magnetometer in a field of  
10 kOe. The data were
taken upon cooling from 700 K at which the samples were kept for about
5 minutes.

\begin{figure}[htb]
    \includegraphics[height=5.5cm]{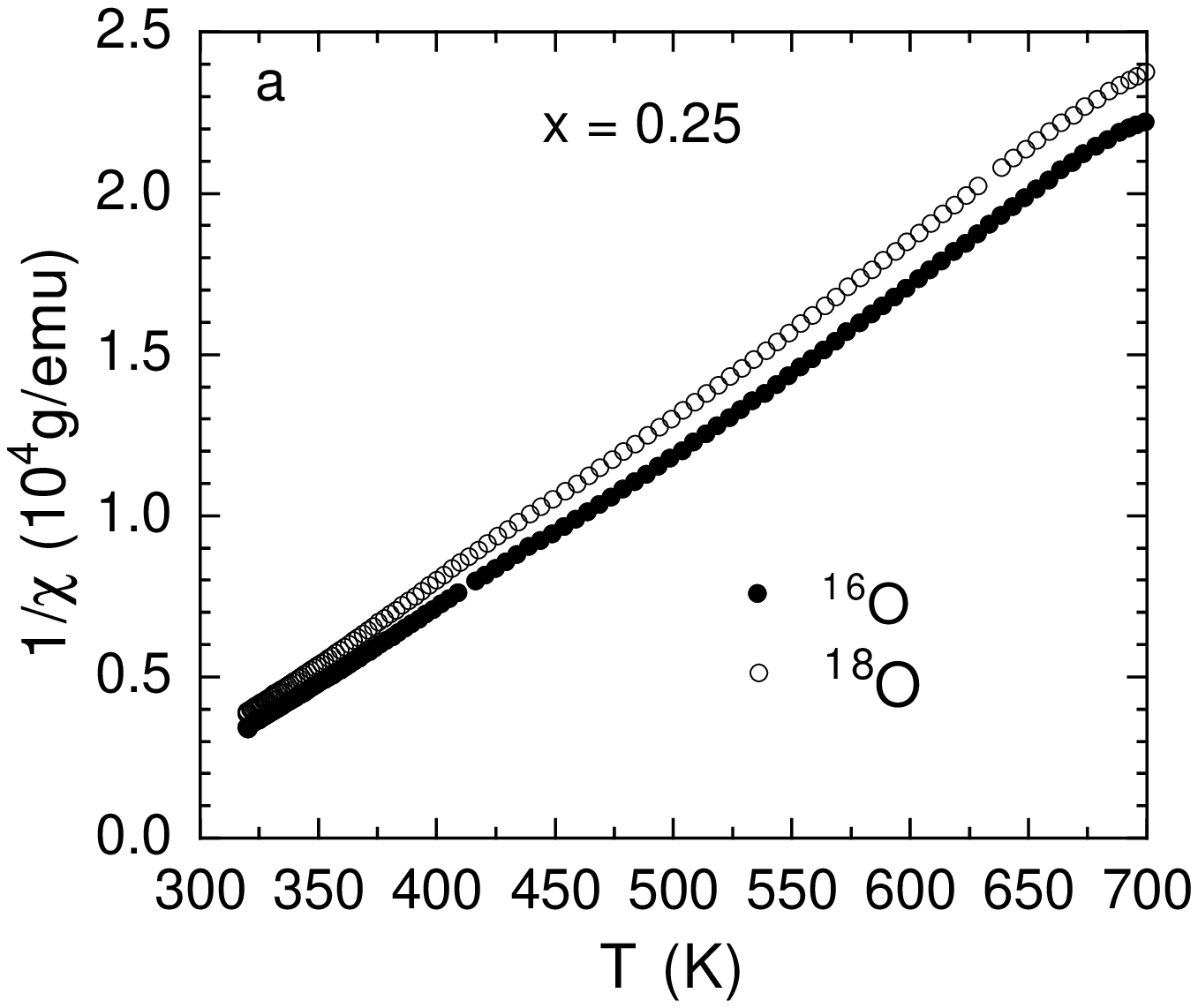}
	 \includegraphics[height=5.5cm]{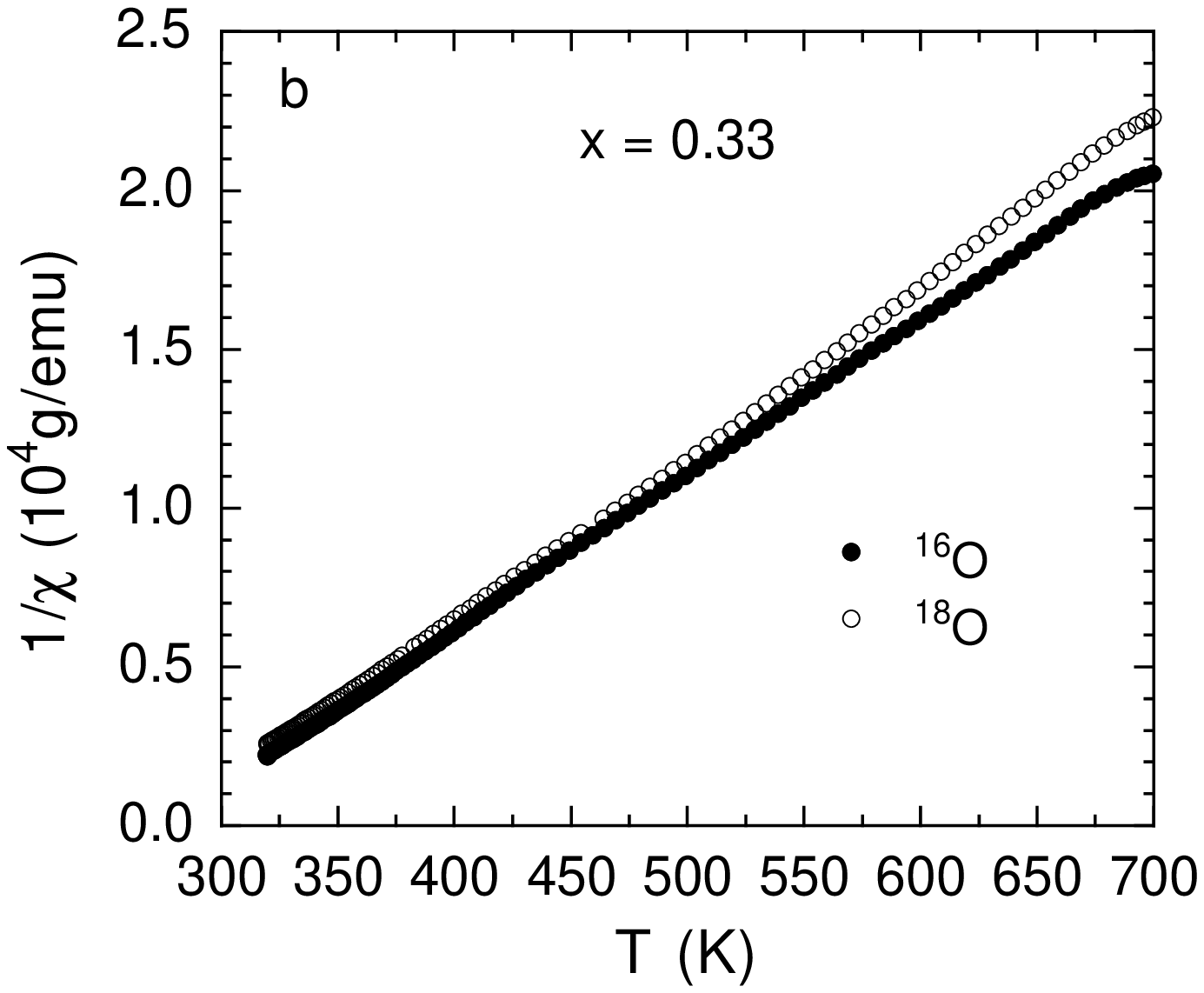}
 \caption[~]{Temperature dependencies of the inverse
susceptibilities of the $^{16}$O and $^{18}$O samples
of La$_{1-x}$Ca$_{x}$MnO$_{3}$ with $x$ = 0.25 (a)
and $x$ = 0.33 (b) in the high-temperature region (320-700 K).  }
\end{figure}

In Figure 1, we show temperature dependencies of the normalized
magnetizations (normalized to the magnetization
well below T$_{C}$) for the $^{16}$O and $^{18}$O samples
of La$_{1-x}$Ca$_{x}$MnO$_{3}$ with $x$ = 0.25 
and = 0.33.  The oxygen-isotope shift of T$_{C}$ was determined from the differences
between the midpoint temperatures on the transition curves of the $^{16}$O and
$^{18}$O
samples.  There are substantial oxygen-isotope 
shifts of T$_{C}$ for both compositions, as indicated in the figures.

Figure 2 shows temperature dependencies of the inverse
susceptibilities of the $^{16}$O and $^{18}$O samples
of La$_{1-x}$Ca$_{x}$MnO$_{3}$ with $x$ = 0.25 (a)
and $x$ = 0.33 (b) in the high-temperature region (320-700 K). It is apparent 
that the inverse
susceptibility depends on the oxygen-isotope mass in the whole
temperature range studied. At higher temperatures, the oxygen-isotope 
effect even becomes larger. In the
paramagnetic state of the ferromagnetic manganites, the susceptibility
$\chi (T)$
should follow the Curie-Weiss law: $\chi (T) = C/(T - \theta)$,  
where $\theta$ is proportional to the ferromagnetic exchange energy
$J$ within the mean-field theory. Plotting the inverse susceptibility
$1/\chi
(T)$ against $T$ will show a straight line if $\theta$ is independent
of temperature. 

\begin{figure}[htb]
    \includegraphics[height=6.2cm]{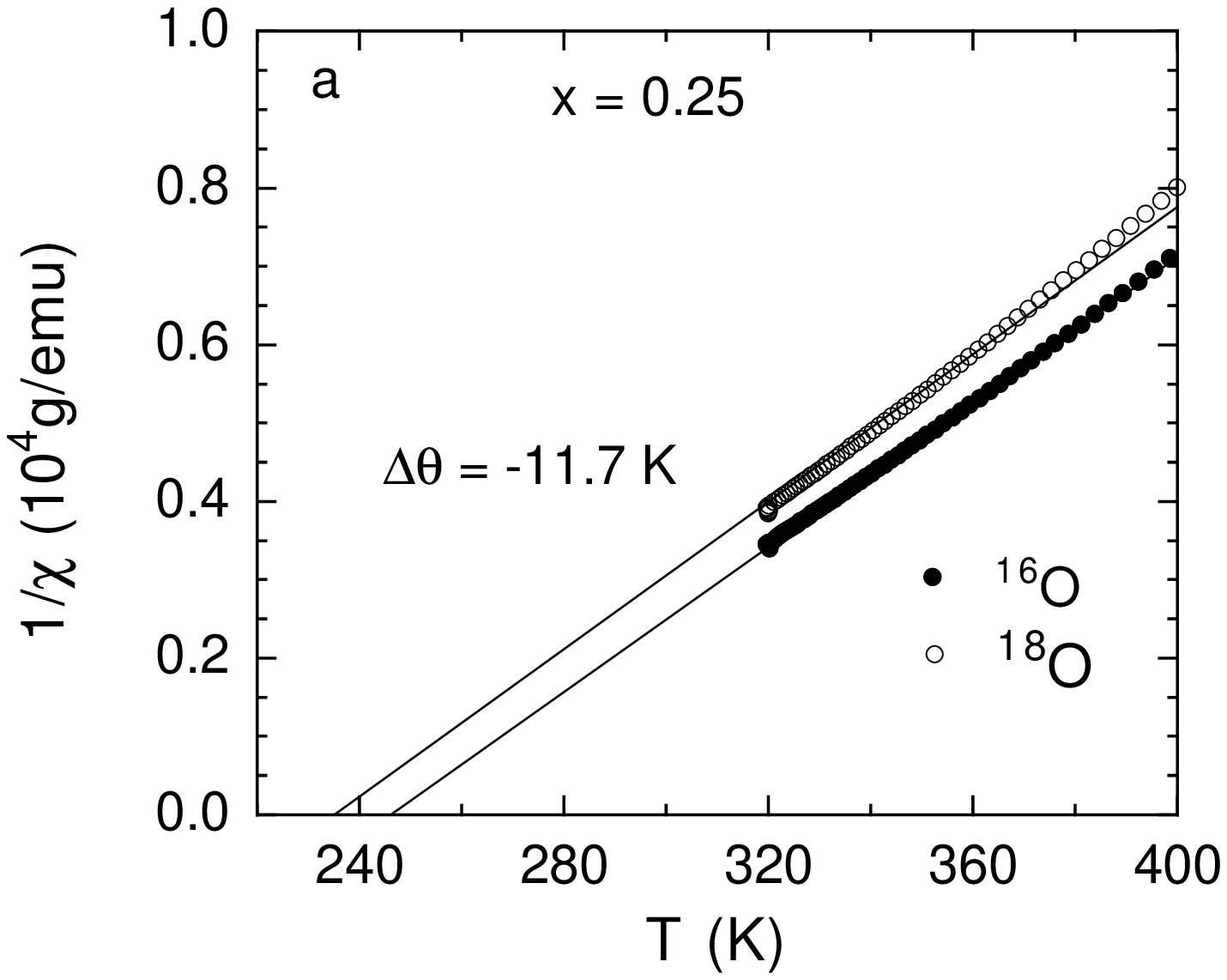}
	 \includegraphics[height=6.2cm]{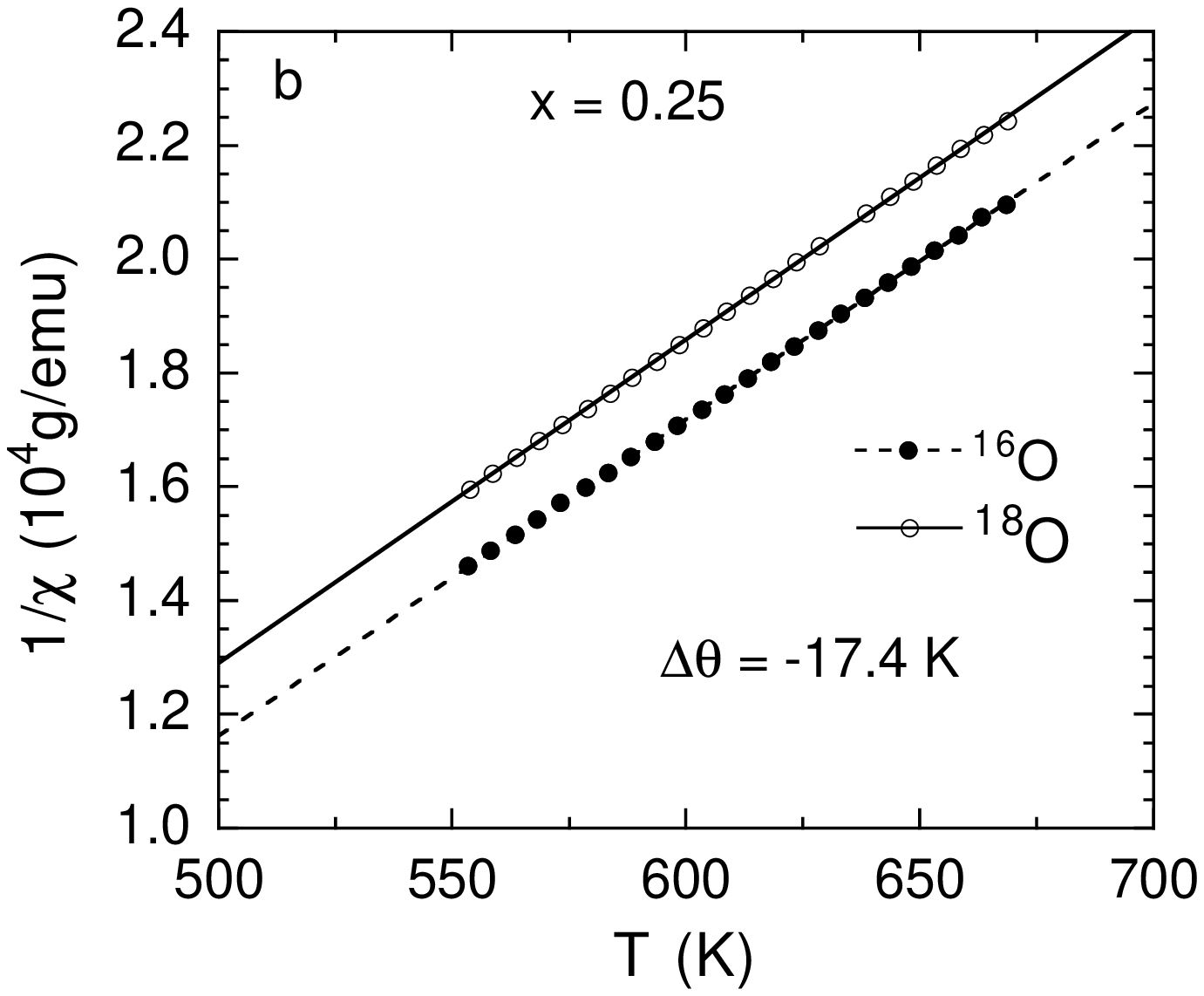}
 \caption[~]{Temperature dependencies of the inverse
susceptibilities of the $^{16}$O and $^{18}$O samples
of La$_{1-x}$Ca$_{x}$MnO$_{3}$ with $x$ = 0.25 in the temperature
region of 320-400K (a) and in the temperature range of 550-670 K (b).  }
\end{figure}

In Figure 3, we show the inverse
susceptibilities of the $^{16}$O and $^{18}$O samples
of La$_{1-x}$Ca$_{x}$MnO$_{3}$ with $x$ = 0.25 in the temperature
region of 320-400K (a) and in the temperature range of 550-670 K (b).
The intercept of the
linear line to the temperature axis  determines the $\theta$ value. In the temperature
region of 320-360K, the $\theta$ value is lowered from 247.4 K to 235.7 
K upon replacing $^{16}$O by $^{18}$O. The oxygen-isotope shift in the $\theta$
value is 11.7 K, which is slightly smaller than  the oxygen isotope
shift of the Curie temperature (14.5 K). On the other hand, in the temperature
region of 550-670 K, the $\theta$ value is lowered from 290.8 K to 273 K
upon replacing $^{16}$O by $^{18}$O. The oxygen-isotope shift in the $\theta$
value is 17.4 K, which is slightly larger than  the oxygen-isotope
shift of the Curie temperature.

Figure 4 shows the inverse
susceptibilities of the $^{16}$O and $^{18}$O samples
of La$_{1-x}$Ca$_{x}$MnO$_{3}$ with $x$ = 0.33 in the temperature
region of 320-400K (a) and in the temperature range of 600-680 K (b). In the temperature
region of 320-360K, the $\theta$ value is lowered from 274.8 K to 267.0 
K upon replacing $^{16}$O by $^{18}$O. The oxygen-isotope shift in the $\theta$
value is 7.8 K, which is slightly smaller than  the oxygen-isotope
shift of the Curie temperature (9.7 K). In the temperature
region of 600-680 K, the $\theta$ value is lowered from 301.4 K to 289.2 K
upon replacing $^{16}$O by $^{18}$O. The oxygen-isotope shift in the $\theta$
value is 12.2 K, which is slightly larger than  the oxygen-isotope
shift of the Curie temperature.

\begin{figure}[htb]
    \includegraphics[height=5.5cm]{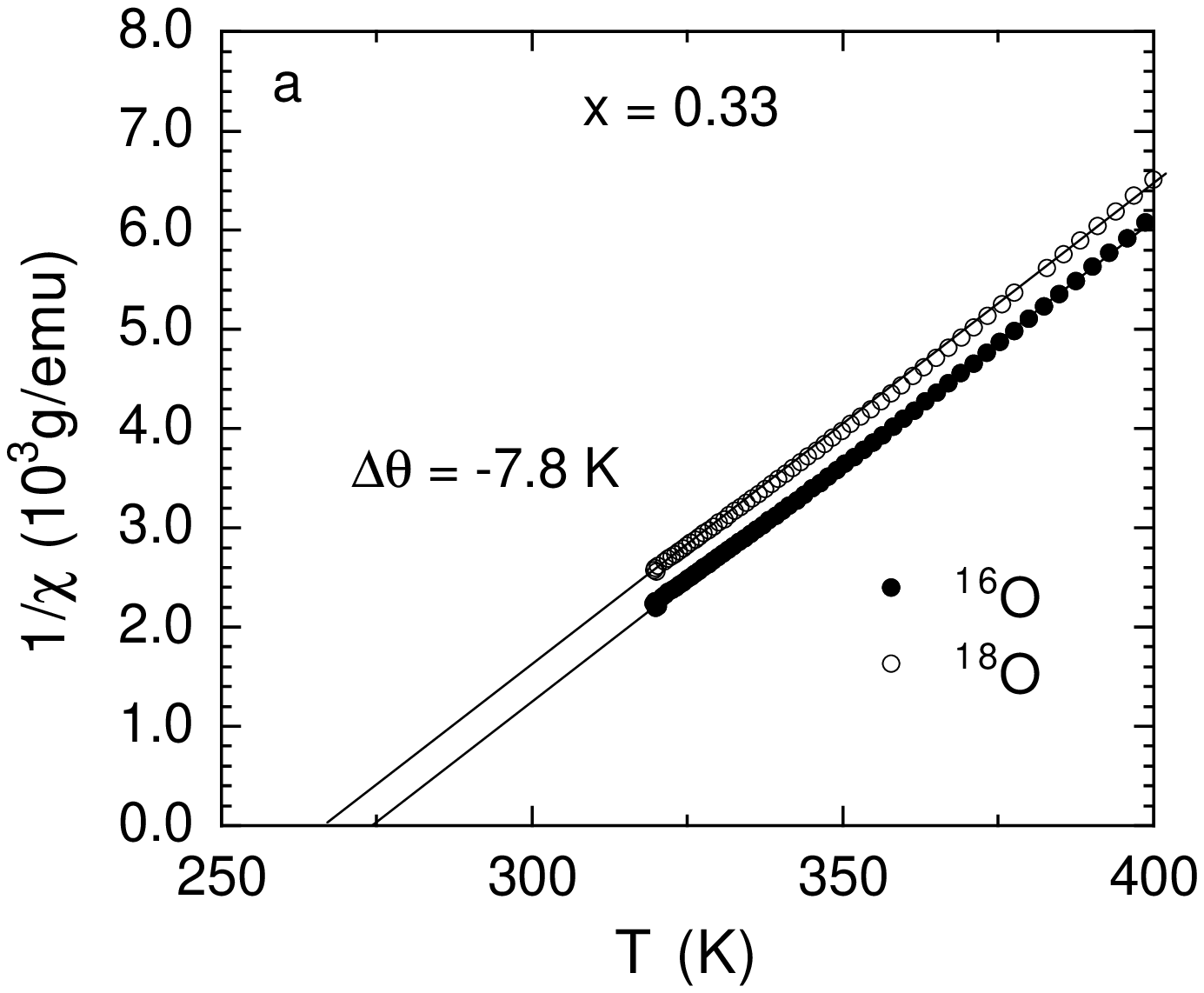}
	 \includegraphics[height=5.5cm]{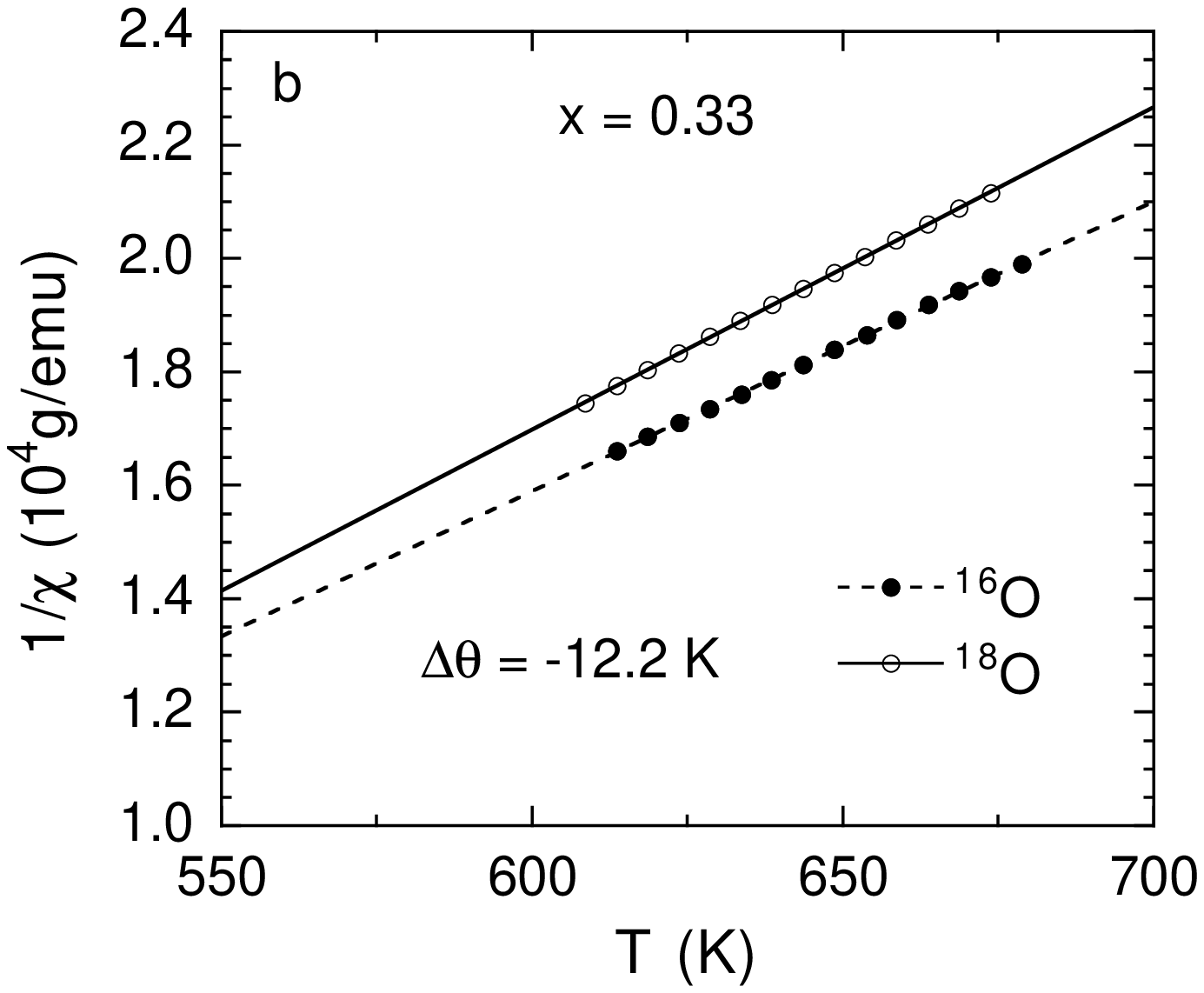}
 \caption[~]{Temperature dependencies of the inverse
susceptibilities of the $^{16}$O and $^{18}$O samples
of La$_{1-x}$Ca$_{x}$MnO$_{3}$ with $x$ = 0.33 in the temperature
region of 320-400 K (a) and in the temperature range of 600-680 K (b).  }
\end{figure}

In Figure 5, we plot $\theta$ value as a function of temperature for
the $^{16}$O  and $^{18}$O samples
of La$_{0.75}$Ca$_{0.25}$MnO$_{3}$. The $\theta$ values
are obtained from fitting the inverse susceptibility data within
a 25-30 K temperature
interval by the Curie-Weiss law. It is apparent that the temperature
dependence of the $\theta$ value is not monotonic; there is a local
minimum at about 450 K, which may be related to the formation of ferromagnetic clusters
\cite{Teresa}. Moreover, the oxygen-isotope shift of the $\theta$
value decreases with temperature 
up to about 400 K and then increases again with temperature above
about 400 K. The much smaller isotope effect on $J$ around 400 K appears to be related to the
much smaller isotope efect on the thermoelectric  power around 400 K
(Ref.~\cite{ZhaoPRBHT}). 
The temperature dependencies of these isotope effects below 400 K are consistent with 
the AB model. However, the increase of the isotope effect above 400 K is not expected from the AB
model which predicts a negligible isotope
effect on $J$ at high temperatures when bipolaronic
carriers is diminishing. Therefore, the original AB model, which is based on the 
pure $p$-$d$ exchange and
the formation of oxygen-hole bipolarons, is inconsistent with the
high-temperature isotope effect. 

\begin{figure}[htb]
    \includegraphics[height=6cm]{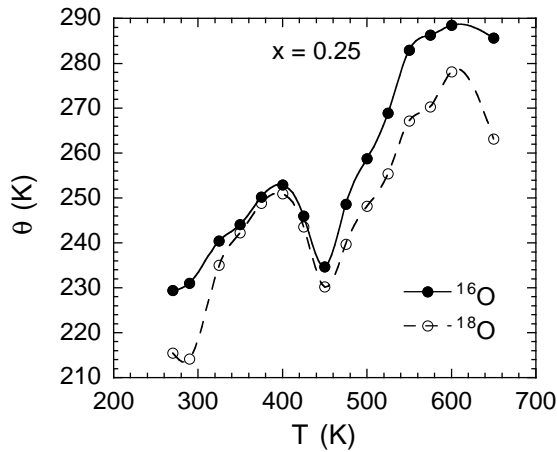}
 \caption[~]{Temperature dependencies of the  $\theta$ values for
the $^{16}$O  and $^{18}$O samples
of La$_{0.75}$Ca$_{0.25}$MnO$_{3}$. The $\theta$ value at each
temperature
is obtained from fitting the inverse susceptibility data within an
$\sim$30 K temperature
interval by the Curie-Weiss law. }
\end{figure}

Since the observed oxygen-isotope effects on the intrinsic
electrical transport properties are consistent with the formation of
bipolarons \cite{ZhaoPRBHT}, we should not discard the AB model based on
the formation of localized bipolarons in the paramagnetic state.
Photoemission spectroscopy \cite{Sat} indicates that about 23$\%$ of doped holes 
reside on Mn 3d orbitals while 77$\%$ of doped holes sit on oxygen
orbitals in the low doping range. In the heavily doped range, doped holes
should have more 3d character so that neither the pure $p$-$d$ exchange
nor the pure double-exchange is responsible for 
the ferromagnetism. In order to agree
with the present
isotope-effect results, one should assume that the ferromagnetic exchange
energy in these samples also depends significantly on the effective mass of polarons.
This is possible when the ferromagnetism is also caused by
double-exchange of doped holes in the Mn-3d orbitals (about 30$\%$) although the pure
double-exchange cannot explain the ferromagnetism \cite{ZhaoDE}.

In summary, we have measured high-temperature magnetic
susceptibilities of the
oxygen-isotope exchanged La$_{1-x}$Ca$_{x}$MnO$_{3+y}$. The
data show that the ferromagnetic exchange energy $J$ depends strongly on the
oxygen-isotope mass. The temperature dependence of the isotope effect 
is very unusual, which cannot be explained by existing 
theories of the colossal magnetoresistance effect for doped manganites.
The present results thus provide essential constraints on the physics
of manganites.

 ~\\
 ~\\
\noindent 
{\bf 
Acknowledgment:} This research is supported by a Cottrell Science Award from Research 
Corporation. We thank the Palmdale Institute of Technology for the use of the VSM and 
Lockheed Martin Aeronautics for the cryogens.\\
~\\
$^{*}$ gzhao2@calstatela.edu
\bibliographystyle{prsty}

\end{document}